\documentclass[twocolumn,aps,showpacs]{revtex4}
\usepackage{epsfig}

\begin{document}
 
\title{Anticipated synchronization: a metaphorical linear view}
\author{Oscar Calvo$^1$, Dante R. Chialvo$^{1,2,3}$, V\'\i ctor M. 
Egu\'\i luz$^3$, Claudio Mirasso$^1$ and Ra\'ul Toral$^{1,3}$}

\affiliation{$^1$Departament de F\'{\i}sica, Universitat de les Illes 
Balears, E-07122 Palma de Mallorca, Spain \\
$^2$Department of Physiology, Northwestern University, Chicago,
Illinois, 60611, USA \\
$^3$Instituto Mediterr\'aneo de Estudios Avanzados IMEDEA (CSIC-UIB), Ed. Mateu Orfila, Campus UIB,
E-07122 Palma de Mallorca, Spain}
\date{\today}

\begin{abstract}
We study the regime of anticipated synchronization recently described 
on a number of dynamical systems including chaotic and noisy ones. We 
use simple linear caricatures to show the minimal setups able to 
reproduce the basic facts described.
\end{abstract}

\maketitle

\section{Introduction}

The coupling of dynamical systems can lead to the synchronization of their
outputs \cite{PRK01}. Recently, attention has been given to the a peculiar
phenomenon introduced by Voss \cite{V00,V01a,V01b,V02} where one system
synchronizes not with the present state but with some future state of another
system. This regime, called {\sl anticipated synchronization}, has been
demonstrated theoretically and numerically in disparate dynamical systems
including discrete or continuous, chaotic and noisy excitable, and
quasi-periodic ones  \cite{M01,MZ01,HMM02,CCMMT02,TMMCC02}. Experimental
results have considered either electronic circuit implementations of the
dynamical equations  \cite{V02,CCMMT02} or lasers running in the chaotic regime
both in one-directional \cite{LTDASL02} and bidirectional coupling
\cite{SSSS01,APL98,HFEMM01,BVPMTG02}.

From the viewpoint of an external observer, the dynamics of anticipated
synchronization can be seen as if one system is forecasting the state of the
other. It is unclear, however, under which general conditions a given dynamical system would or not exhibit such dynamics. The phenomenon in itself is rather counterintuitive because, depending on the setting, mixes notions of synchrony and order with the dynamics of chaos and disorder. Furthermore, it is not immediately apparent whether or not nonlinearities are essential for the process. This paper is dedicated to answer these concerns, by analyzing the minimal setup able to exhibit the more fundamental aspects of this phenomena. 
 
The paper offers three perspectives covering from the simplest scenario to the more complex ones. Section~II is a brief overview of anticipated synchronization
as described in recent work. Section~III discusses a toy model describing the trivial situation of two ``particles'' moving with uniform trajectory in
which one attempts to closely follow the other using the scheme described in
the anticipated synchronization literature. These particles can be seen
as an special case of the dynamical system presented in Section~IV,
where the case of two coupled maps is analyzed from a dynamical
systems perspective and the condition for anticipated synchronization
derived analytically. The same system is revisited in Section~V, but
now from a Control Systems point of view, both for continuous and
discrete cases. The paper closes by listing the most relevant
conditions one would expect to see in all cases of anticipated
synchronization.
 
\section{Overview of anticipated synchronization.} 

Two different schemes have been proposed in order to achieve
anticipated synchronization \cite{V00}. Both schemes use, in a way or
another, delay lines which allow forecasting of a {\sl master}
trajectory by a {\sl slave} identical system. 

The first scheme uses the technique of {\sl complete replacement}. It considers a dynamical system, ${\bf x}(t)$ (known as the {\sl master system}) whose dynamics involves a delayed feedback of the form:
\begin{equation}
\label{cr1}
\dot {\bf x}(t) = -\alpha {\bf x}(t)+{\bf f}({\bf x}(t-\tau)),
\end{equation}
being $\alpha>0$ a constant, and ${\bf f}$ a given, generally non--linear,
function.  The dynamics of the {\sl slave system}, ${\bf y}(t)$, is obtained by a similar equation in which the delay term has been completely replaced by the
master system. Namely, the evolution equation for ${\bf y}(t)$ is:
\begin{equation}
\label{cr2}
\dot {\bf y}(t) = -\alpha {\bf y}(t)+{\bf f}({\bf x}(t))
\end{equation}
It is easy to see that the manifold ${\bf y}(t)={\bf x}(t+\tau)$, in which the slave anticipates by a time $\tau$ the actual output of the master, is a stable solution of the dynamical equations (\ref{cr1}-\ref{cr2}). This follows readily from the (exact) evolution equation $\dot {\bf \Delta} (t) =-\alpha {\bf \Delta}(t)$ for the delayed difference ${\bf \Delta} (t)={\bf y}(t)-{\bf x}(t+\tau)$. This result is independent of both the function ${\bf f}({\bf x})$ under consideration, or the arbitrarily large delay time $\tau$. The structural stability of the asymptotic solution has been demonstrated by implementation in electronic circuits \cite{V02}. Of course, the result is more remarkable when the dynamics of the master has a high degree of unpredictability, {{\it i.e.}} by choosing  ${\bf f}({\bf x})$ and $\tau$ such that the dynamics of the master is chaotic. In this scheme, and beyond the mathematical result, the anticipation mechanism can be understood from the fact that the dynamics of the master at time $t$ influences its own dynamics at a time $\tau$ later, whereas it enters the dynamics of the slave immediately at time $t$. In other words, it is as if the slave system ${\bf y}(t)$ has {\sl anticipated knowledge} of an essential part of the dynamics of the master system ${\bf x}(t)$.

The second scheme is one that includes only delay in the slave dynamics \cite{V00}. This is defined by the following dynamical equations for the master and slave systems:
\begin{equation}
\label{scheme1}
\begin{array}{rcl}
\dot {\bf x}(t)& =& {\bf f}({\bf x}(t))\\
\dot {\bf y}(t) & = & {\bf f}({\bf y}(t))+{\bf K}[{\bf x}(t)-{\bf y}(t-\tau)],
\end{array}
\end{equation}
where ${\bf f}({\bf x}(t))$ is an arbitrary function and ${\bf K}$ is a coupling strength matrix. It is easy to show that the anticipated synchronization manifold \mbox{${\bf y}(t)={\bf x}(t+\tau)$} is again a solution of this set of equations. At variance with the case of complete replacement, its stability can only be fulfilled in a limited range of parameters for $\tau$ and $\bf K$. Despite this restriction, however, we believe that this method of anticipated synchronization deserves more attention than that of complete replacement because the anticipation time $\tau$ is included as an external parameter and does not influence the dynamics of the master system. Therefore, in principle, any dynamical system can be predicted by using this scheme. Again, the anticipation is more relevant when the dynamics of the master is unpredictable by other simple means.

Recently an extension of the scheme was introduced to include
anticipation in non--autonomous systems \cite{CCMMT02,TMMCC02}.
Specifically, it has been considered the following set of equations:
\begin{equation}
\label{scheme2}
\begin{array}{rcl}
\dot {\bf x}(t)& =& {\bf f}({\bf x}(t))+{\bf I}(t)\\
\dot {\bf y}(t) & = & {\bf f}({\bf y}(t))+{\bf I}(t)+{\bf K}[{\bf x}(t)-{\bf y}(t-\tau)],
\end{array}
\end{equation}
where the new term ${\bf I}(t)$ represents an external input acting on
both, master and slave, systems. Although the anticipated manifold
\mbox{${\bf y}(t)={\bf x}(t+\tau)$} is no longer an {\sl exact}
solution of the previous equations (except in the case of a periodic
forcing ${\bf I}(t+\tau)={\bf I}(t)$), it has been shown that several
features of the dynamics of the master can indeed be predicted by the
slave. For instance, it is possible to predict the peaks fired by an
excitable system subjected to a random external forcing
\cite{CCMMT02,TMMCC02}. We stress that in this case the random forcing
induces peak firing at uncorrelated and unpredictable times.

In this second scheme, Eqs.~(\ref{scheme1}-\ref{scheme2}), the actual mechanism
leading to synchronization is much more elusive and, despite the wide variety
of work, it is still unclear which are the relevant conditions and requirements
for two dynamical systems to exhibit this type of anticipated synchronization.
For instance, one would like to understand whether or not nonlinear aspects of
the dynamics are needed for the systems to exhibit anticipated synchronization.
In the same line to what extent anticipation can be arbitrarily long is of
relevance for practical purposes.

\section{Two particles following each other} 

The intention in this section is to undress the anticipated synchronization models of irrelevant aspects, to be able to identify the essential mechanisms at
play. The goal is to have a system homologous to the more general ones
described in the anticipated synchronization literature. We first look at the simplest linear one, 
thus we choose a dynamical system of two particles moving
uniformly in  a one-dimensional space. We consider a particle following
a uniform motion
\begin{equation}
\dot x(t) = v~,
\label{m}
\end{equation}
where $x(t)$ is the position of the particle at time $t$ and $v$ its velocity.
Following the scheme in anticipated synchronization to forecast (and ``synchronize'' with) the
position of the master particle we consider another (``slave'')
particle whose trajectory is given by
\begin{equation}
\dot y(t) = v + K(x(t) - y(t-\tau))~,
\label{s}
\end{equation}
where $K$ is the strength of the coupling
between the master particle (Eq.~(\ref{m})) and the slave (Eq.
(\ref{s})). To achieve the anticipation, the solution $y(t) =
x(t+\tau)$ has to be a stable solution. The stability can be studied
analyzing the behavior of the delayed difference $\Delta(t) = y(t) -
x(t+\tau)$. We obtain
\begin{equation}
\dot \Delta(t) = - K \Delta(t-\tau)~.
\end{equation}
The condition for local stability of the solution $\Delta(t) = 0$ ($y(t) = 
x(t+ \tau)$) is given
by \cite{Hayes50,Glass88}
\begin{equation}
K \tau < \frac{\pi}{2}~.
\label{taomax}
\end{equation}
This indicates that in order to observe an anticipated solution of the
slave particle within this scheme, the product of the coupling with the
anticipation time $\tau$ has to be smaller than a certain value. Larger values produce
an over-correction and the slave is unable to anticipate. This
expression qualitatively reproduces the numerical results obtained by
several researchers.

It is interesting to note that if the two systems are not identical,
e.g., the velocities are different, then
\begin{equation}
\dot y(t) = v^\prime + K(x(t) - y(t-\tau))~.
\end{equation}
then we obtain the same stability condition. However the anticipation
is not perfect in the sense that the slave anticipates with a constant
mismatch \makebox{$\displaystyle y(t) = x(t + \tau ) + \frac{v^\prime - v}{K}$}. The mismatch
decreases with increasing coupling constant $K$.

More generally one could obtain a stability equation of the from
\begin{equation}
\dot \Delta(t) = A \Delta(t) - K \Delta(t-\tau)~,
\end{equation}
leading to the conditions  \cite{Hayes50,Glass88}
\begin{equation}
K < |A|
\end{equation} or,
\begin{equation}
K > |A| \qquad {\rm and} \qquad \tau < \frac{\cos^{-1} (A/K)}{(K^2 - A^2)^{1/2}}~, 
\end{equation}
where the principal value is taken ($0 \le \cos^{-1} (A/K) \le \pi$).

\section{ A dynamical systems perspective: Two coupled maps}

Let us consider the same two particles of the previous section but
now moving with the more general dynamics given by the
following coupled maps:
\begin{equation}
\begin{array}{lcl} 
x_{k+1}  =  \alpha x_k+a ~~~~~~~~ & & {\rm mod}(m)\cr
y_{k+1}  =  \alpha y_k+a+ \gamma(x_k-y_{k-n}) & & {\rm mod} (m),
\end{array}
\label{t}
\end{equation}
where $\alpha>0$ and $a$ are constants. If $\alpha>1$ the $x_k$ form a chaotic map, whereas $\alpha=1$ leads to a quasi-periodic map and for $\alpha<1$ the map converges to \mbox{$x_k=a/(1-\alpha)$}. The necessary $n+2$ initial conditions are the set $(x_0,y_{-n},y_{-n+1},\cdots, y_{-1},y_0)$.

We are interested in the possibility of the coupled maps leading to anticipated
synchronization, {\it i.e. } in having
\begin{equation}
y_k=x_{n+k}~.
\end{equation}
as an asymptotic solution for the maps.
To this end, we consider the map satisfied by the delayed difference  $\delta_k=x_{n+k}-y_k$:
\begin{equation}
\delta_{k+1}=\alpha\delta_k -\gamma\delta_{k-n}.
\end{equation}
An analysis complementary to ours can be found in  \cite{HMM02}. The ansatz $\delta_k=\lambda^k$ leads to
\begin{equation}
\delta_k=\sum_{i=1}^{n+1}C_i \lambda_i^k\hspace{0.5truecm}{\rm mod} (m)~,
\end{equation}
The (complex) constants $C_1,\dots,C_{n+1}$ are determined by the
initial conditions. The $\lambda_i$, $i=1,\dots,n+1$ are the solutions of the
polynomial equation:
\begin{equation}
\label{roots}
\lambda^n (\lambda-\alpha)+\gamma=0~.
\end{equation}
A necessary and sufficient condition for anticipated synchronization to hold and
to be asymptotically stable is that $\displaystyle \lim_{k\to \infty}\delta_k=0$,
or equivalently $|\lambda_i|<1,\,\forall  i=1,\dots,n+1$. We analyze now the
range of validity of this condition as a function of the parameters $\alpha$, $n$
and $\gamma$. It turns out that anticipated synchronization occurs if $\gamma\in(\alpha-1,\gamma_c)$, where the value of $\gamma_c$ depends on $\alpha$ and $n$. We consider three cases:

{\bf 1.-} $\alpha<1$. We treat for completeness this case, although it is the less interesting from the practical point of view, since the maps tend asymptotically to a fix point (in practise, after a finite number of steps). Although a full analytical solution does not seem to be available in this case, it is possible to obtain the asymptotic behavior in the case of large anticipation $n$:
\begin{equation}
\gamma_c \to 1-\alpha +\frac{\pi^2}{2n^2}, 
~~~~~~\alpha<1~,
\end{equation}
Notice that anticipation is possible for arbitrarily large $n$. 
 
{\bf 2.-} $\alpha=1$. In this case, the dynamics is quasi-periodic. The interval of anticipated synchronization can be found analytically as:
\begin{equation}
\label{alfa1}
\gamma_c=2\sin\left(\frac{\pi}{2(2n+1)}\right).
\end{equation}
In the limit $n\to\infty$, $\gamma_c \sim \frac{\pi}{2n}$, the same as the continuous case considered in the previous section. Again, anticipation is possible for arbitrarily large $n$.

\begin{figure}[htbp]
\centering
\psfig{figure=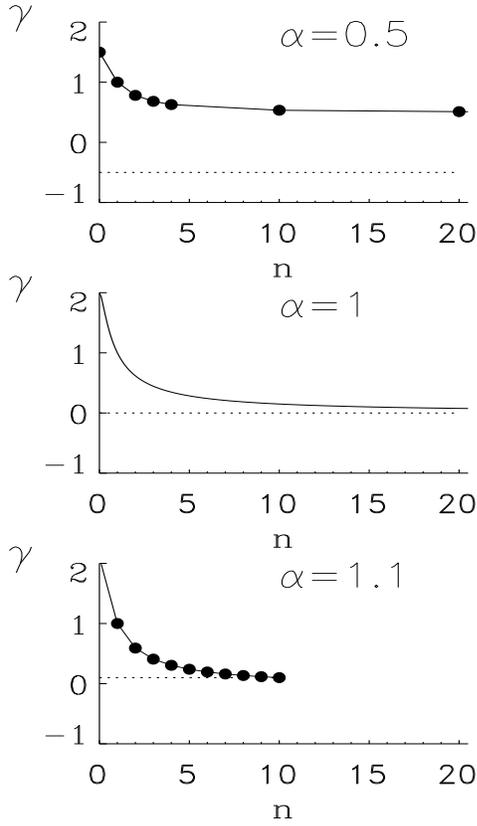,width=7.cm,height=12.0cm,angle=0,clip=true}\caption{\label{figalpha}The anticipated synchronization regime is achieved for values of $\gamma$ and $n$ in between the upper (solid) line and the lower (dashed) line. For $\alpha=1$ the solid line is the analytical result Eq. (\ref{alfa1}), whereas for $\alpha=0.5$ and $\alpha=1.1$ the dots have been computed numerically and the solid line is a guide to the eye.}
\end{figure}

{\bf 3.-} $\alpha>1$. This is the more interesting case since the maps are chaotic. It turns out that the
condition $|\lambda_i|<1$ is only satisfied for $\alpha <
1+\frac{1}{n}$.  Alternatively, for fixed $\alpha >1$ the maximum
anticipation time is
\begin{equation}
n_{max}= \left[\frac{1}{\alpha-1}\right]~~~~~~\alpha>1~,
\end{equation}
where $[x]$ denotes the integer part of $x$. Anticipated synchronization is found for
$\gamma\in(\alpha-1,\gamma_c)$ but  only for $n< n_{max}$. In this case
of $\alpha>1$ we have not been able to find any analytical approximation or asymptotic expression and the values of $\gamma_c$ need  to be computed numerically. The previous results are summarized in Fig.\ref{figalpha} 

\begin{figure}[htbp]
\centering
\psfig{figure=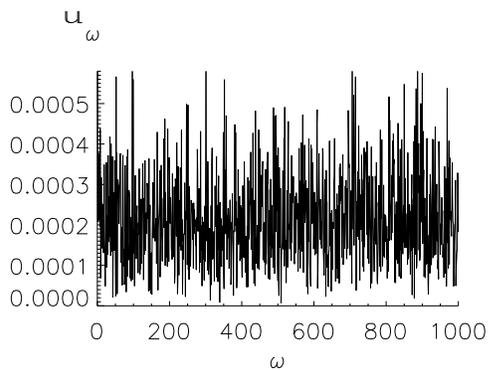,width=7.5cm,clip=true}
\caption{\label{figsk}Power spectrum $u_{\omega}$ (arbitrary units) of the series of $u_k$ pulses generated by Eq.~\ref{t} with parameters $\alpha=1.1$, $a=0.1$ and $m=1$. }
\end{figure}

Let us now analyze the predictability of the map, taking $\alpha=1.1$, $n=6$,
for which the synchronization limits are $\gamma\in(0.1,0.19395)$. First, for
visualization purposes, we transform the output of the master map into a series
of random spikes by defining a new ``firing"  sequence $u_k$ as $u_k=1$ if
$x_k>u$ and $u_k=0$  if $x_k<u$, with $u$ a convenient threshold value.
Similarly for the slave map, we define $v_k=1$ if $y_k>u$ and $v_k=0$ if
$y_k<u$.  In the case of $\alpha=1.1$ the timing of the pulses does not have any
regularity. As shown in Fig.~\ref{figsk}, the power spectrum of the $u_k$
signal is flat, showing the absence of any preferred time scale, the series is
highly unpredictable.

\begin{figure}[htbp]
\centering
\psfig{figure=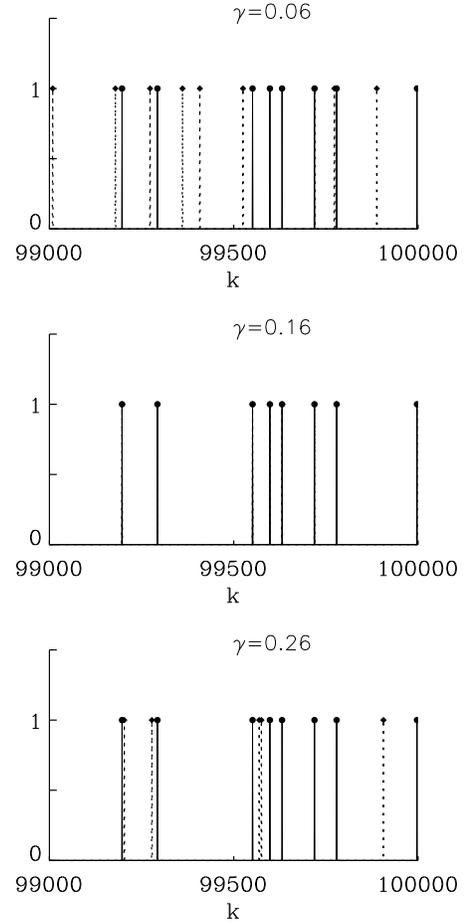,width=7.cm,height=14.cm,angle=0,clip=true}
\caption{\label{figukvk} Pulses $u_k$ (solid lines and circles) and $v_{k-n}$  (dashed lines and diamonds), generated by the iteration of Eq.~\ref{t}, with $\alpha=1.1$, $a=0.1$, $n=6$, for three values of $\gamma$ . Notice that anticipated synchronization (coincidence of the $u_k$ and $v_{k-n}$ pulses) occurs only for the intermediate value of $\gamma$ (middle panel).}
\end{figure}

In Fig.~\ref{figukvk} we show that, in accordance with the previous
analysis, each pulse of the master system $u_k$ is anticipated $n=6$
units of time by a pulse in the slave system $v_{k-n}$ for a coupling
$\gamma=0.16\in(0.1,0.19395)$ (middle panel). For too large or too small values of the
coupling $\gamma$, the synchronization is lost. The existence of a
minimum and maximum value for the coupling in order to have anticipated
synchronization, which appears here as a property of simple linear maps,
has also been observed in more complex chaotic \cite{V00} and excitable
non-autonomous systems \cite{TMMCC02}. These maps can be seen as well 
as a case of integrate and fire dynamics, often used as caricatures of neural systems. 

\section{A Control Systems Perspective} 

From an engineer viewpoint, the slave dynamical system following a master,
shown in the previous section, is seen as a cascade of control system blocks.
In this section anticipated synchronization is analyzed from this Control
System perspective. In both, the continuous and discrete time systems, ${\bf
x}(t)$ will be seen as the output of a first order open loop system with no
feedback, and the slave ${\bf y} (t)$ (controlled signal) represents the output
of an identical system but with delayed time feedback and driven by the master
system ${\bf x}(t)$ (the reference signal). With enough loop gain in the
feedback loop, the control will act as a servo mechanism which minimizes the
error between the reference (master ${\bf x}(t)$) and its own delayed output
${\bf y} (t-\tau)$. Hence, the internal variable ${\bf y} (t)$ will be a {\it
prediction} of ${\bf x}(t)$.

\subsection{Two particles following each other seen as a servo mechanism}

The example of the two particles following each other described by
equations  (\ref{m}) and (\ref{s}) can be seem as a servo mechanism
where the position of the controlled particle $y(t-\tau )$ is compared
with a reference signal $x(t)$. The integrator nature of the loop
provides enough gain so that in steady state the error will vanish and
the two signals will match. 

Equation (\ref{m}) represents a pure integrator to input $v$.
Denoting by $X(s)$ the Laplace transform of $x(t)$ and assuming zero
initial conditions for $x(t=0)$, the Laplace transform of
$\dot x(t)$ will be $sX(s)$. Also, since the
constant velocity $v$ is represented as an external step input, its
Laplace transform $V(s)$ will be equal to  $v/s$. Thus we can write 
\begin{equation}
X(s)=\frac{v}{s^2}~.  \label{ms1}
\end{equation}
This represents a transfer function, in the Laplace domain, with a pole
at the origin of the $s$ plane. If $v$ is constant the output is a
ramp $x(t)=vt$.

Repeating the same process for Eq.~(\ref{s}) we can write 
\begin{equation}
sY(s)=\frac{v}{s} + K(X(s) - e^{-\tau s}Y(s))~,  \label{ss1}
\end{equation}
from where we obtain
\begin{equation}
Y(s)=\frac{v/s}{(s+Ke^{-\tau s})}+\frac{KX(s)}{(s+Ke^{-\tau s})}~.
\label{ss3}
\end{equation}
Applying the inverse Laplace Transform to Eq.~(\ref{ss3}) we can obtain
the temporal evolution of the output for any given input. The dynamics
of the error in the servo mechanism can also be obtain by subtracting $X(s)-Y(s).$

\begin{figure}[htbp]
\centering
\psfig{figure=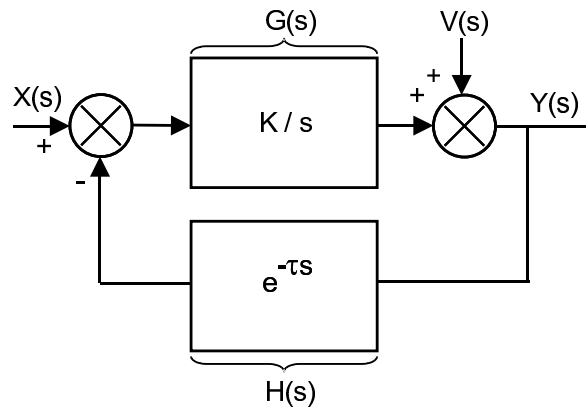,width=3.0 in,angle=0,clip=true}
\caption{\label{figGs}Block diagram  of  the continuous-time  servo mechanism system (Eq. (\ref{ss1})--(\ref{ss4})) represented by a transfer function in the Laplace domain. }
\end{figure}

The input/output relation, ({\em transfer function} in control system terms), between $X(s)$ and $Y(s)$ is given by the ratio of these two functions (considering zero external input):
\begin{equation}
G_{cl}(s)=\frac{Y(s)}{X(s)}=\frac{K/s}{1+ \frac{K}{s} e^{-\tau s}}~.
\label{ss4}
\end{equation}
Equations (\ref{ss1})--(\ref{ss4}) can be represented in a block
diagram, as in Fig.~\ref{figGs}, with direct gain $G(s)=\frac{K}{s}$
and feedback gain $H(s)=e^{-\tau s}$. In more general terms, the closed
loop transfer function is given by
\begin{equation}
G_{cl}(s)=\frac{Y(s)}{X(s)}=\frac{G(s)}{1+G(s)H(s)}~.
\end{equation}
It follows that if the loop
gain $G(s)H(s) \gg 1$, the closed loop transfer function will be given by the
feedback term $G_{cl}\simeq (H(s))^{-1}$. For our example this means that
$Y(s)\simeq X(s)e^{\tau s}$, and applying the inverse Laplace transform
$y(t)\simeq x(t+\tau )$. This was the expected result that proves that the
delayed position of the second particle, after some transient, will follow the
position of the first one. Likewise, $y(t)$ will be moving ahead of $x(t)$ in a
predictive manner.

To perform a stability analysis, the roots of the denominator
$s + K e^{-\tau s}=0$ must be found. If we replace $s$ by
$i\omega $ a frequency analysis can be obtained. This is usually done
drawing real and imaginary parts of the transfer function with the
Nyquist diagram or using the Bode plots, representing module and phase
separately. The term $G(i\omega ) \simeq K/i\omega $ has a gain whose
module decreases with $\omega $ and becomes unity for $\omega =K$, and a
constant phase equal to $-\pi /2$. The transfer
function of the delay $H(i\omega )$ has a constant module equal to
one 
\begin{equation}
\left| H(i\omega )\right| =\left| e^{-i\omega\tau }\right| =1~,
\end{equation}
and a phase that decreases with $w$
\begin{equation}
\phi (\omega )={\rm arg}[e^{-i\omega\tau}]=-\omega \tau ~.
\end{equation}

According to the standard stability criteria for linear systems the
module of the gain of $G(\omega )H(\omega )$ must be smaller than $1$
when the phase crosses the $\pi $ boundary (meaning the
denominator of $G_{cl}(s)$ is zero). Substituting $\omega =K$ the
total phase contribution $(\tau K+\pi /2)$ at the unity gain point
should be smaller than $\pi $, recovering Eq.~(\ref{taomax}) 
\begin{equation}
\tau <\frac{\pi }{2K}~.
\end{equation}

\subsection{Discrete System}

A discrete system similar to the coupled maps described by Eq.~(\ref{t})
(but without the $\mathrm{mod}(m)$ constrain) can be seen, in engineering terms, as a discrete time control systems written as two difference equations: one for the
master $x_k$ and the other for the slave $y_k$.
The equations (\ref{t}) without the $\mathrm{mod}(m)$ constrain can be
rewritten for time $k$ 
\begin{eqnarray}
x_{k+1} &=&\alpha x_{k}+a  \label{mdk1} \\
y_{k+1} &=&\alpha y_{k}+a+\gamma (x_{k}-y_{k-n})  \label{sdk1}
\end{eqnarray}

\begin{figure}[htbp]
\centering
\psfig{figure=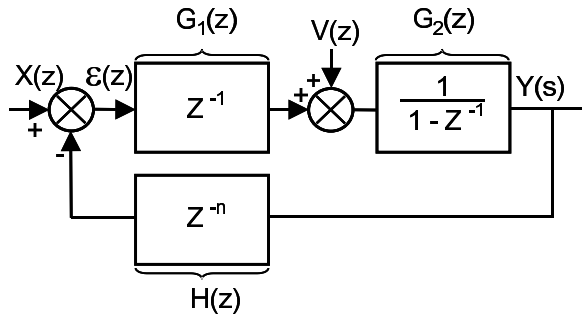,width=3.0 in,angle=0,clip=true 
}
\caption{\label{figgGz}Block diagram of the discrete-time servo mechanism system Eq. (\ref{eq6}) represented by a transfer function in the Z-transform domain.}
\end{figure} 

To solve Eqs.~(\ref{mdk1})--(\ref{sdk1}), we will use the $Z$-transform
\cite{Franklin90}. If $Z\{x_{k}\}$ is the $Z$-transform of $x_k$
denoted as $X(z)$ and $Z\{x_{k-1}\}$ is the $Z$-transform of $x_{k-1}$
obtained by multiplying $X(z)$ by $z^{-1}$, applying these rules to
both equations, we obtain
\begin{eqnarray}
X(z)&=&\frac{Z\{a\}}{1-\alpha z^{-1}}	\\
Y(z)&=&\frac{Z\{a\} +\gamma z^{-1}X(z)}
{1-\alpha z^{-1}-\gamma z^{-(n+1)}}~,  \label{eq5}
\end{eqnarray}
where $Z$-transform of $a$, $Z\{a\}$, depends on the type of
input $a$. For our case, $a$ is constant and
\begin{equation}
Z(a)=\frac{az}{z-a}~.
\end{equation}

As in the continuous system, we are interested in the transfer function between
$X(z)$ and  $Y(z)$.
Any feedback system can be viewed as a direct transfer function $G(z)$ and a
feedback block $H(z)$ whose output will be subtracted from the reference, to
obtain the error signal. In our case
\begin{equation}
Y(z)=	\frac{G_{1}(z)G_{2}(z) X(z) + G_{2}(z) Z(a)}
{1+G_{1}(z)G_{2}(z)H(z)}~,
\label{eq6}
\end{equation}
where $G_{1}(z)=\gamma z^{-1}$, $G_{2}(z)=(1-\alpha
z^{-1})^{-1}$, and $H(z)=z^{-n}$.
In the engineering literature, these equations are often represented using block diagrams, as shown in Fig.~\ref{figgGz}.
The closed loop transfer function in this case is given by
\begin{equation}
G_{cl}(z)=\frac{Y(z)}{X(z)}
=\frac{\gamma z^{-1}}{1-\alpha z^{-1}+z^{-n}z^{-1}}~.
\end{equation}
Using the same arguments as in the previous subsection, if the loop
gain $G_{1}(z)G_{2}(z)H(z) \gg 1$ we can approximate
$G_{cl}\simeq (H(z))^{-1}$. This yields to
$Y(z)\simeq z^{n}X(z)$ indicating that the output $y_k$ predicts
$x_k$ by $n$ sampling periods.

To analyze the stability we analyze the conditions when the denominator of
Eq.~(\ref{eq6}) equals zero.
Thus to ensure stability the roots of the denominator must lie inside the circle
given by 
\begin{equation}
z^{n}(z-\alpha )=-\gamma ~,
\end{equation}
recovering Eq.~(\ref{roots}).

\section{Conclusions}

As stated at the outset, the purpose of this paper is to undress the
models of anticipated synchronization to be able to see the essentials
of this intriguing dynamics, which in summary are:

i) The phenomenon of anticipated synchronization in itself does not rely on nonlinear properties of the dynamics, indeed, as shown in Sect. IV nonlinearities make the anticipation harder or impossible. 

ii) Some aspects of the problem are naturally understood by looking it from a Control Systems approach, where the delayed term are seen as an ``error signal'' in a servo mechanism control loop (Sect. V). 

iii) As shown by the derivation using simple stability criteria (Sect. III and IV), the boundaries of the anticipated synchronization (i.e., the fundamental
diagram in Fig.~1 of coupling strength versus the delay) are expected to be universal for this kind of systems.

Overall, these results account for the necessary qualitative features two systems must have in order to exhibit anticipated synchronization. By clarifying under which conditions the phenomenon would be observable, some questions will naturally arise opening new application of these ideas. From the arguments discussed here it seems that ``synchronizing in advance'' and ``controlling the future state'' can be equivalent objectives describing the aim of the process of anticipated synchronization. In that sense it might be worth to study possible connections with systems that learn to master such objectives, as for instance the task of a neural net adapting to capture a flying object.

{\bf Acknowledgements} DRC thanks the hospitality and support of the Departamento de F{\'\i}sica,  Universitat de les Illes Balears, Palma de Mallorca.  This work is supported
by the Spanish Ministerio de Ciencia y Tecnolog{\'\i}a and FEDER projects BMF2001-0341-C02-01, BFM2002-04474-C02-01, BFM2000-1108 and by NINDS (DRC).

\end{document}